\documentclass[]{spie}  

\usepackage{amsmath,amsfonts,amssymb}
\usepackage{graphicx}
\usepackage[colorlinks=true, allcolors=blue]{hyperref}
\usepackage{float}
\usepackage[utf8]{inputenc}

\title{NMF-based GPU accelerated coronagraphy pipeline}

\author[a]{Sai Krishanth P.M.}
\author[a]{Ewan S. Douglas}
\author[a]{Justin Hom}
\author[a]{Ramya M. Anche}
\author[b]{John Debes}
\author[c]{Isabel Rebollido}
\author[d,e]{Bin B. Ren}
\affil[a]{Steward Observatory,  The University of Arizona, 933 North Cherry Avenue, Tucson, AZ 85721, USA}
\affil[b]{Space telescope science institute, 3700 San Martin Drive, Baltimore, MD 21218, USA}
\affil[c]{Centro de Astrobiología, Instituto Nacional de Técnica Aeroespacial, Ctra de Torrejón a Ajalvir, km 4, 28850 Torrejón de Ardoz, Madrid Spain}
\affil[d]{Universit\'{e} C\^{o}te d'Azur, Observatoire de la C\^{o}te d'Azur, CNRS, Laboratoire Lagrange, Bd de l'Observatoire, CS 34229, 06304 Nice cedex 4, France}
\affil[e]{Universit\'{e} Grenoble Alpes, CNRS, Institut de Plan\'{e}tologie et d'Astrophysique (IPAG), F-38000 Grenoble, France}

\authorinfo{Further author information: (Send correspondence to S.K.P.M.)\\S.K.P.M.: E-mail: saikrishanth@arizona.edu}

\begin{document} 
\maketitle

\begin{abstract}
We present a generalized Non-negative factorization (NMF)-based data reduction pipeline for circumstellar disk and exoplanet detection. By using an adaptable pre-processing routine that applies algorithmic masks and corrections to improper data, we are able to easily offload the computationally-intensive NMF algorithm to a graphics processing unit (GPU), significantly increasing computational efficiency. NMF has been shown to better preserve disk structural features compared to other post-processing approaches and has demonstrated improvements in the analysis of archival data. The adaptive pre-processing routine of this pipeline, which automatically aligns and applies image corrections to the raw data, is shown to significantly improve chromatic halo suppression. Utilizing \textit{HST}-STIS and \textit{JWST}-MIRI coronagraphic datasets, we demonstrate a factor of five increase in real-time computational efficiency by using GPUs to perform NMF compared to using CPUs. Additionally, we demonstrate the usefulness of higher numbers of NMF components with SNR and contrast improvements, which necessitates the use of a more computationally efficient approach for data reduction.
\end{abstract}


\section{INTRODUCTION}
\label{sec:intro}  

Direct imaging of debris disks reveals the physical properties and morphology of dust scattering and emission around evolving planetary systems\cite{2018ARA&A..56..541H}. The complex extended structures of disks seen in imaging surveys can reveal dynamical interactions with planets, interactions with binaries, and self-stirring from the disk\cite{2018ARA&A..56..541H}. This can help constrain disk and planet formation theories in a way that indirect detections cannot. Additionally, direct imaging allows us to study populations of planets not accessible with indirect methods such as radial velocity measurements and photometric temporal variations. However, the existence of speckle noise, optical aberrations, and other instrumental artifacts in imaging data makes post-processing necessary. The basis of many high-contrast imaging post-processing algorithms is to use a reference (either a different object or different parts of the same object) to model the starlight present in an image and subtract it to reveal the astronomical signal we are interested in. 

One high-contrast imaging algorithm, non-negative matrix factorization (NMF), is used to construct a non-orthogonal, non-negative set of basis components from a given reference library. The basis components are used to construct a model PSF that is then removed from a science image. As demonstrated in Ren et al. \cite{2018ApJ...852..104R}, NMF can better preserve disk morphology than other methods such as classical Reference-star Differential Imaging (RDI\cite{1984Sci...226.1421S}), Angular Differential Imaging (ADI\cite{2006ApJ...641..556M}), Karhunen-Lo\`eve Image Projection (KLIP\cite{2012ApJ...755L..28S}), and Locally Optimized Combination of Images (LOCI\cite{2007ApJ...660..770L}). Additionally, most of these methods rely on forward modeling to accurately extract the disk signal. Since most forward modeling approaches rely on prior assumptions about the morphology of the disk, this is not desired. NMF, on the other hand, does not require forward modeling and can detect fainter disk structures. It has been demonstrated that higher numbers of NMF components can better capture quasi-static noise properties\cite{2018ApJ...852..104R}. NMF has been used previously to image the debris disk around HD 32297 in Duch\^{e}ne et al.\cite{2017AAS...22914605R}, and for the detection of a faint arc around HD 129590 in Olofsson et al.\cite{2023A&A...674A..84O}. It has also been used for the first dynamical detection of a companion driving a spiral arm in HD 100453 in Xie et al.\cite{2023A&A...675L...1X}.

NMF, however, relies largely on matrix multiplications, and since matrix multiplications have a complexity of O(n$^3$), utilizing a higher number of components in large surveys quickly becomes untenable. Stark et al.\cite{2023ApJ...945..131S} and Ren\cite{2023arXiv230816912R} have previously noted the difficulties of using NMF because of its computationally-intensive nature. The majority of the computation involved in matrix multiplication is easily parallelizable since each row/column pair is operated independently by one thread, and multiple threads can operate in tandem. Using GPUs with substantially larger thread counts than CPUs (often at least an order of magnitude) can, therefore, be quite useful. Additionally, GPU architectures are designed to be massively parallelizable, thus providing an additional increase in the efficiency of code. 

In this paper, we describe our new open-source pipeline that uses NMF for disk and planetary reductions, present run-time improvements relative to CPU-based NMF, demonstrate signal-to-noise ratio (SNR), and contrast improvements by using higher numbers of NMF components. We also present an exoplanet image reduced using our pipeline to begin studying the efficacy of NMF for planetary detections. We provide a brief theoretical overview of the NMF algorithm in section \ref{sec:NMF}, describe data acquisition in section \ref{sec:data acquisition}, go over data reduction procedures in \ref{sec:data reduction}, discuss our findings in \ref{sec:results}, and summarize our findings and detail future work in \ref{sec:summary}.

\section{Non-negative Matrix Factorization} \label{sec:NMF}

The premise of NMF is the factorization of one matrix into two non-negative matrices. Consider the problem:

\begin{equation}
    X = WH
\end{equation}

Where X is the initial matrix we start with, and W and H are the two matrices we wish to factor X into. We must first define a cost function to compare our approximation of X with WH. Because of its ease of implementation, the Frobenius norm is chosen as the cost function. It is defined as follows:

\begin{equation}
    ||V^{1/2} \circ (X-WH)||^2_F
\end{equation}

Where $\circ$ is the Hadamard product, and V is a weighting term that accounts for heteroscedastic uncertainties and missing data. For \textit{Hubble Space Telescope} (\textit{HST}) datasets, this term is derived from the \texttt{ERR} extension (see \S \ref{sec:data reduction}). To iteratively minimize the cost function, we adopt vectorized updated rules provided by Zhu\cite{2016arXiv161206037Z}:

\begin{equation}
\begin{split}
    H \leftarrow H \circ \frac{W^T(V\circ X)}{W^T[V\circ(WH)]}\\
    W \leftarrow W \circ \frac{(V\circ X)H^T}{[V\circ(WH)]H^T}.
\end{split}
\end{equation}

H and W are initialized randomly, and the process is left to continue until the change in gradient from one step to the next reaches a set value or the maximum number of iterations is reached. We use a modified version of the \texttt{nmf\_imaging}\cite{nmfimaging} package in this work, using \texttt{cupy}\cite{cupy_learningsys2017} to offload matrix multiplications to the GPU. 
Additionally, for disk imaging, we also use best factor finding per Ren et al.\cite{2018ApJ...852..104R} (implemented in \texttt{nmf\_imaging}) in the target modeling process to circumvent forward modeling. 

The NMF problem is known to be NP-hard\cite{2007arXiv0708.4149V}, i.e., its solvability and verifiability in polynomial time is unknown. Additionally, the implementation of NMF we use is known to be non-convex and thus does not converge globally\cite{2016arXiv161206037Z}. We are currently studying other implementations to improve both speed and convergence properties.

\section{Data acquisition} \label{sec:data acquisition}

To test the efficacy of our NMF pipeline, we perform PSF subtraction for 5 debris disk systems observed using the Space Telescope Imaging Spectrograph (STIS) instrument on \textit{HST} in coronagraphic imaging mode. All observations were taken with the \textit{WedgeA1.0} occulter position. We retrieved MAST\footnote{https://archive.stsci.edu/ \label{link-mast}} archival data for AU Microscopii, HD 107146, and HD 181327 from HST GO-12228 (PI: G. Schneider; Schneider et al. \cite{2012AAS...21944602S}). Data for 49 Ceti was retrieved from HST GO-15218 (PI: É. Choquet, Ren, et al. \cite{2023A&A...672A.114R}), and data for HD 163296 was retrieved from HST GO-15437 (PI: E. Rich, Rich et al. \cite{2020ApJ...902....4R}). We selected this sample because of the varying stellar and disk properties to test the efficacy of NMF in different observational scenarios.

\begin{table}[ht]
\label{tab:diskproperties}
\begin{center}       
\begin{tabular}{|l|l|l|l|} 
\hline
\rule[-1ex]{0pt}{3.5ex}  Target name & Spectral Type & Magnitude (R) & Inclination (degrees)  \\
\hline
\rule[-1ex]{0pt}{3.5ex}  49 Ceti \cite{2017ApJ...839...86H}   & A1  & 5.6 & 79 \\
\hline
\rule[-1ex]{0pt}{3.5ex}  AU Mic \cite{2023arXiv230802486L}    & M1  & 8.9 & 89.2  \\
\hline
\rule[-1ex]{0pt}{3.5ex}  HD 107146 \cite{2014AJ....148...59S} & G2  & 6.7 & 18  \\
\hline
\rule[-1ex]{0pt}{3.5ex}  HD 163296 \cite{2018ApJ...869L..42H} & A1e & 6.9 & 48  \\
\hline
\rule[-1ex]{0pt}{3.5ex}  HD 181327 \cite{2014AJ....148...59S} & F5  & 7.1 & 32  \\
\hline 
\end{tabular}
\end{center}
\caption{Properties of our debris disk sample.} 
\end{table}

We also apply our pipeline to one imaged planetary system, HIP 65426b, observed in coronagraphic imaging mode using \textit{JWST}-Mid-Infrared Instrument (MIRI). Data from the MAST archive was obtained from ERS-01386 (PI: S. Hinkley, Hinkley et al. \cite{2022PASP..134i5003H}, Carter et al. \cite{2023ApJ...951L..20C}).

\section{Data reduction} \label{sec:data reduction}

\subsection{\textit{HST}-STIS}

First, we acquire the flat field calibration files from the MAST archive. For each target, we acquire all files corresponding to the same occulter position. This gives us access to the \texttt{SCI} and \texttt{ERR} extensions, which are used in the construction of NMF basis components and target modeling (see \S \ref{sec:NMF}), and the \texttt{DQ} extension, which contains flags to perform bad pixel correction. 

\texttt{SCI} and \texttt{ERR} frames with exposure times higher than the median exposure time of a file are discarded, removing saturated frames. The images are aligned with the star at the center using \texttt{radonCenter}\footnote{https://github.com/seawander/centerRadon}. This is done as the default central pixel values are inconsistent for image alignment across many targets. Additionally, some central pixel values needed to be changed manually before center searching because they were off by several hundred pixels, which rendered \texttt{centerRadon} unusable. We then define a square region in every frame, the size of which depends on the extent of the diffraction spikes. The remaining steps of the reduction are performed within this region, as smaller matrix sizes are much faster to process as opposed to the entire image. A 3$\times$3 median filter mask is applied to pixels with \texttt{DQ} values of 16, 256, and 8192 as in \cite{2017SPIE10400E..21R} corresponding to pixels with a dark rate $>5\sigma$ times the median dark level, bad pixels because of a reference file, and pixels with cosmic rays, respectively. A $r^{1/2}$ mask is also applied to correct for the stellar illumination factor\cite{2014ApJ...789...58S}. Finally, both \texttt{SCI} and \texttt{ERR} frames from the flat fielded files are divided by the exposure times for normalization. 

To ensure that only pixels with the intended astronomical signal are used in processing, we construct an algorithmic mask for the central wedge and diffraction spike in STIS image data. Following the approach in Ren et al.\cite{2020ApJ...892...74R}, masked regions are treated as missing data and ignored. Circumstellar signals do not interfere with component construction, assuming that the signal contribution from the masked regions is small.

The target PSFs are then compared to the reference library, and a percentage of the closest frames, depending on the number of desired NMF components, are selected. The closest frames are calculated by minimizing for Euclidean norm as this is the cost function used by NMF. These are then used to construct the component basis. This is done to curtail the chromatic halo caused by stellar color mismatch (see Figure \ref{fig:AUMicimprovements} and \S \ref{sec:pipeline improvements}) and improve the efficiency of basis construction. The NMF components are then used to model each target frame individually, and the best factor-finding procedure is performed. After target modeling, we perform PSF subtraction to extract the disk signal. All NMF-processed, subtracted images are rotated North-up using their position angles with \texttt{pyKLIP}\cite{2015ascl.soft06001W}\footnote{https://pyklip.readthedocs.io/en/latest/pyklip.html?highlight=rotate\#pyklip.klip.rotate}, and then combined using \texttt{ccdproc}\cite{matt_craig_2017_1069648} to increase signal to noise ratio (SNR) and coverage in the sky.

\begin{figure}[ht]
\centering
\includegraphics[width=0.71\textwidth]{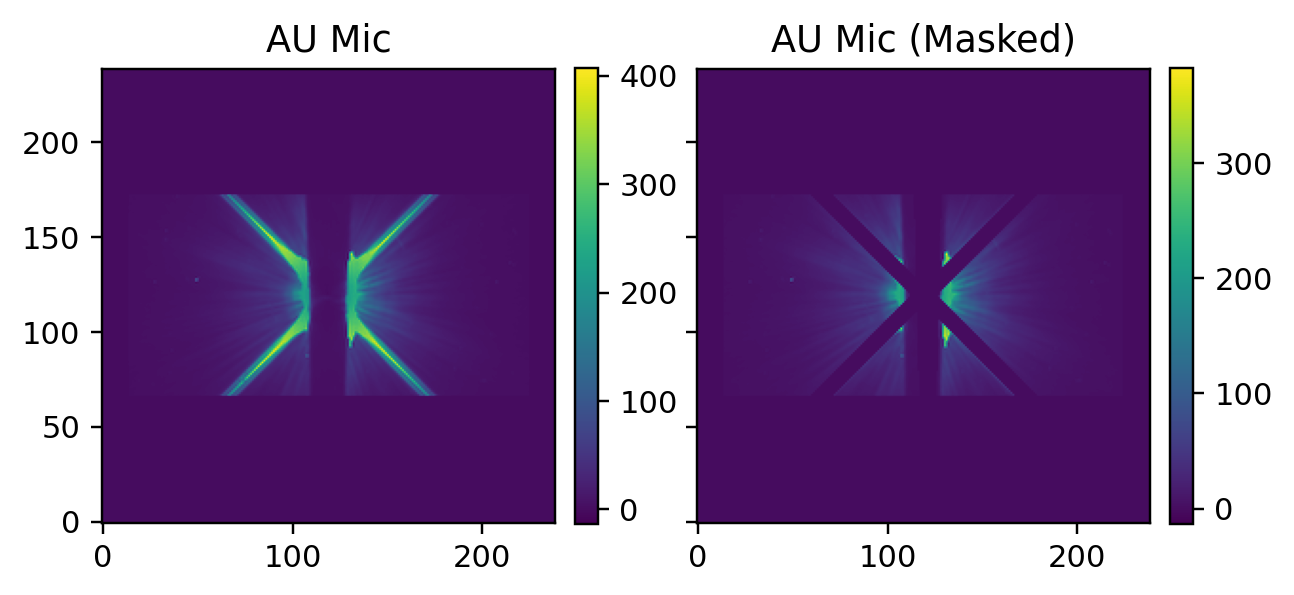}
\caption{Example of \textit{HST}-STIS \textit{WedgeA1.0} frame of AU Microscopii. \textit{Left:} Frame obtained after the preprocessing routine. \textit{Right:} Same as the left but with a mask applied to cover the coronagraphic wedge and diffraction spikes. Masked frames are used in the NMF reduction.}
\end{figure}

\subsection{\textit{JWST}-MIRI}

We acquired raw and uncalibrated (\texttt{*uncal.fits}) files containing HIP 65426 and HIP 68245 data from the MAST archive. Following Carter et al. \cite{2023ApJ...951L..20C}, we processed these files to Stage 1 (\texttt{*rateints.fits}) using a jump detection threshold of 8 in \texttt{SpaceKLIP}\cite{2022SPIE12180E..3NK}. Stage 1 files were further processed to Stage 2 (\texttt{*calints.fits}) using the same pixel cleaning procedures as Carter et al. \cite{2023ApJ...951L..20C} in \texttt{SpaceKLIP}\cite{2022SPIE12180E..3NK}. 

The \texttt{SCI} and \texttt{ERR} frames were then extracted from the stage 2 files, and a square region of 80$\times$80 pixels centered around \texttt{CRPIX1} and \texttt{CRPIX2} was chosen for the NMF reduction. The rest of the procedure is very similar to the processing of STIS data; the reference frames were used to construct the NMF component bases, the bases were used to model the target PSFs, and PSF subtraction was performed. The frames were rotated with \texttt{pyKLIP}\cite{2015ascl.soft06001W} using their position angles and stacked using \texttt{ccdproc}\cite{matt_craig_2017_1069648} to improve SNR.

\begin{figure}[ht]
\centering
\includegraphics[width=0.71\textwidth]{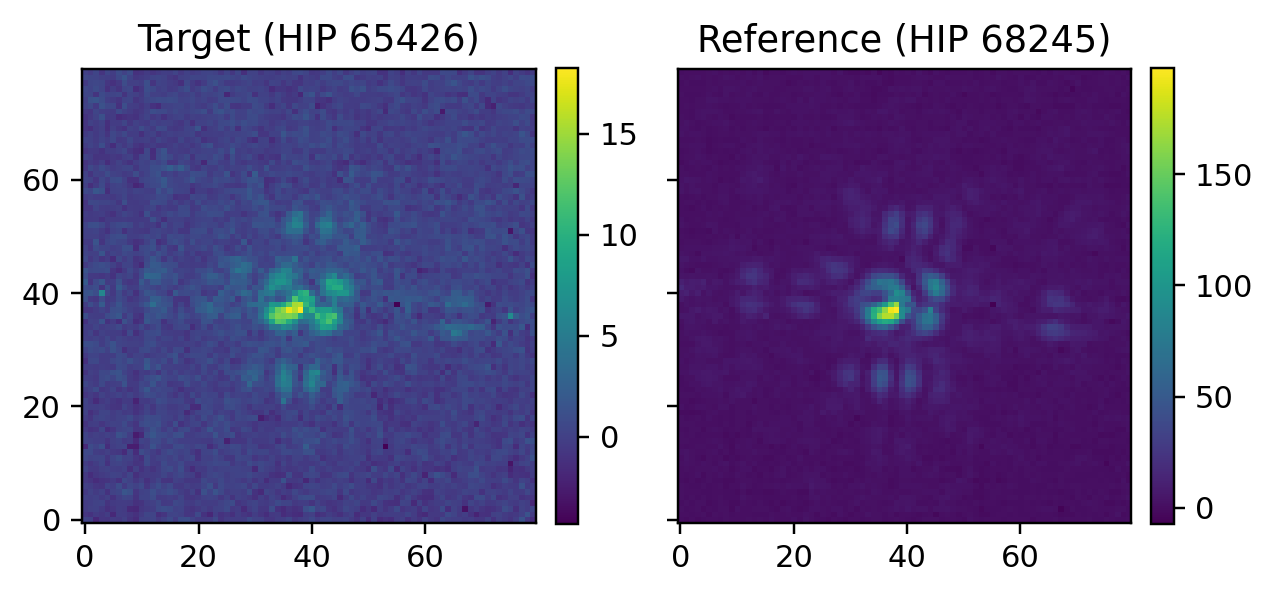}
\caption{Example of \textit{JWST}-MIRI 4QPM F1140C target and reference frames used in the planetary reduction. These were obtained after the preprocessing routine and used in the NMF reduction.} \label{fig:JWSTreduction}
\end{figure}

\section{Results}\label{sec:results}

\subsection{STIS Disk Images}

\begin{figure}[H]
\centering
\includegraphics[width=0.5\textwidth]{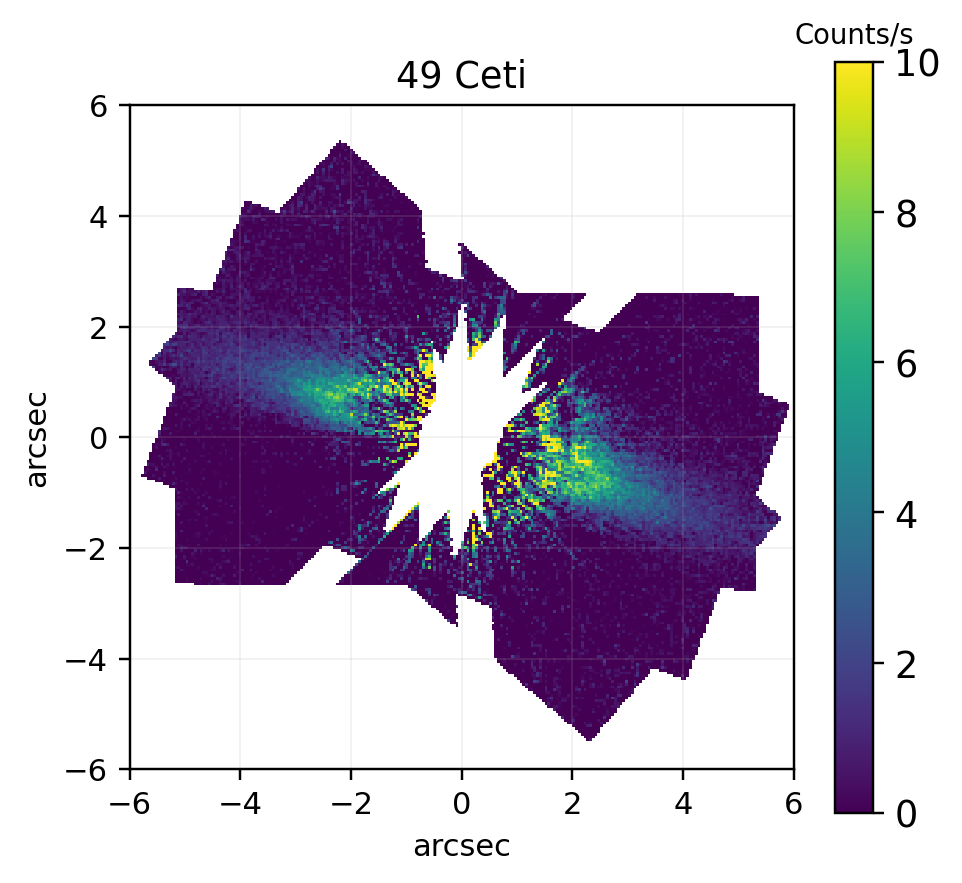}\hfill
\includegraphics[width=0.5\textwidth]{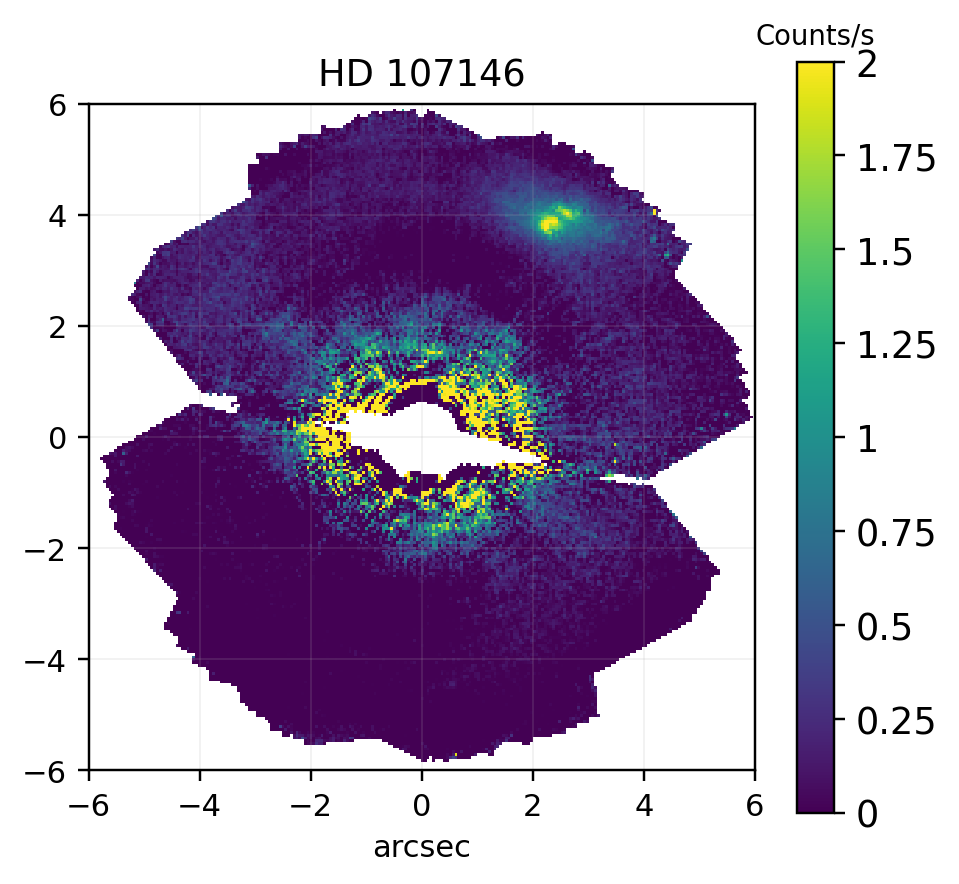}
\includegraphics[width=0.5\textwidth]{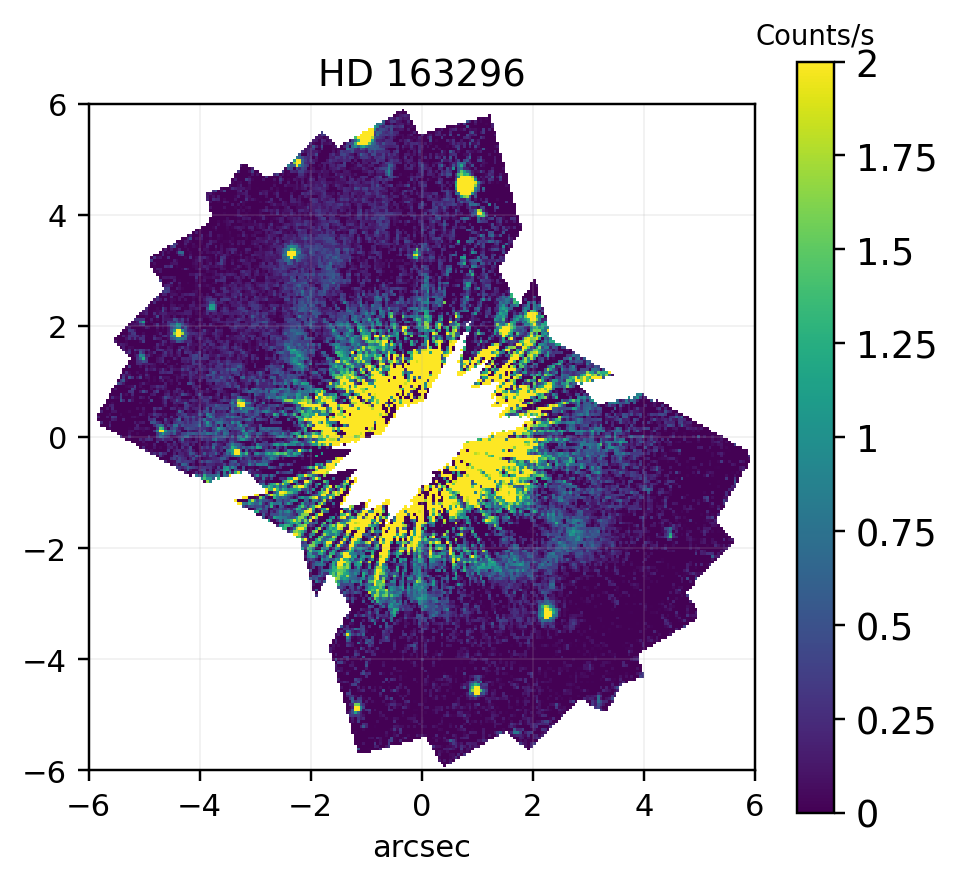}\hfill
\includegraphics[width=0.5\textwidth]{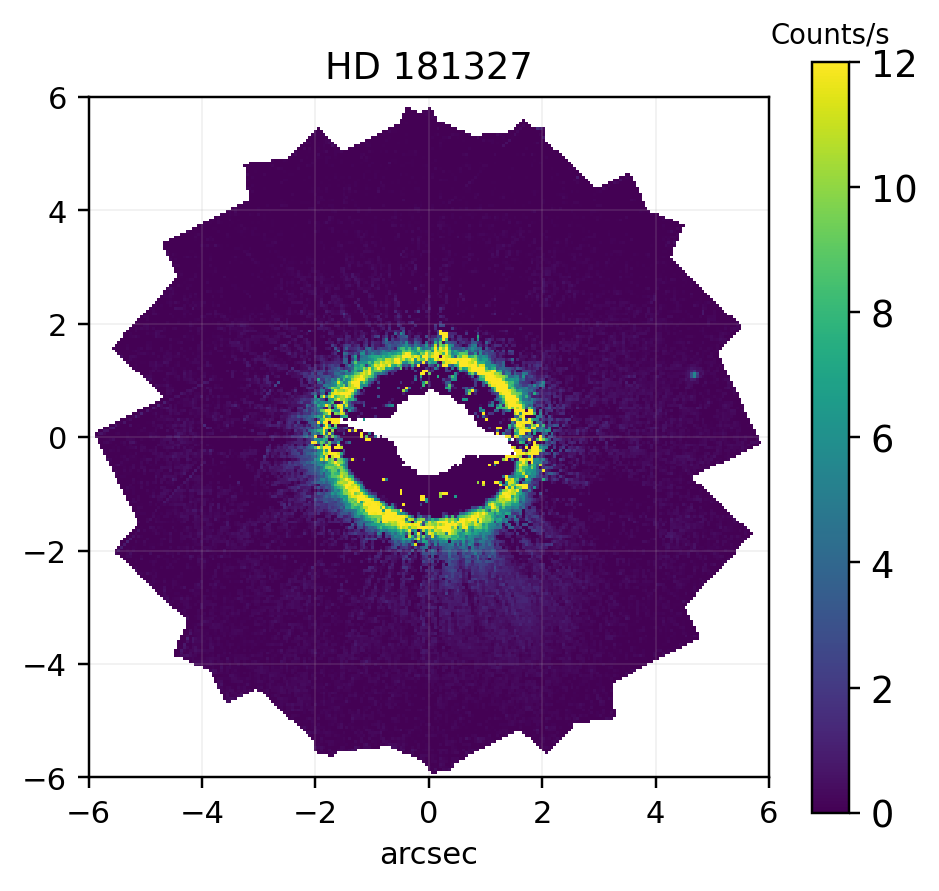}
\caption{A selection of disks reduced using our pipeline. The top left shows 49 Ceti, the Top right shows HD 107146, the Bottom left shows HD 163296, and the Bottom right shows HD 181327. All of these reductions used 90 NMF components to model the target PSF.} \label{fig:NMF disk reductions}
\end{figure}

We show five different disk reductions in this paper, four in Figure \ref{fig:NMF disk reductions} and one in Figure \ref{fig:AUMicimprovements}. The disks shown have different properties, the most important of which are summarized in Table \ref{tab:diskproperties}.

\subsection{HIP 65426b}

For our reduction of HIP 65426b, we only use HIP 68245 frames as our PSF library as the archive of MIRI observations is not as extensive as STIS for our reference to select the closest frames across multiple PSF stars. Analogous to Carter et al. \cite{2023ApJ...951L..20C}, we use the maximum number of NMF components (limited by the size of our library), 81, to model the target before performing PSF subtraction.

\begin{figure}[H]
\centering
\includegraphics[width=0.9\textwidth]{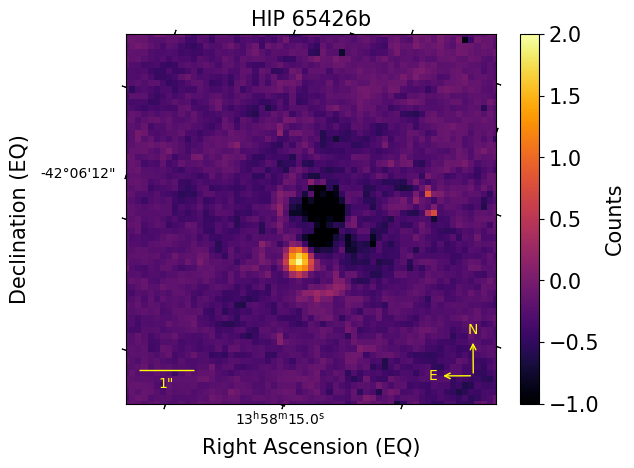}
\caption{NMF-reduced image of HIP 65426b from \textit{JWST}-MIRI 4QPM F1140C data. This figure was generated using 81 NMF components constructed from a PSF library of HIP 68245 frames. We have attempted to reproduce the same figure size as Carter et al. \cite{2023ApJ...951L..20C}.}
\end{figure}

\subsection{Pipeline improvements} \label{sec:pipeline improvements}

\begin{figure}[ht]
\centering
\includegraphics[width=0.5\textwidth]{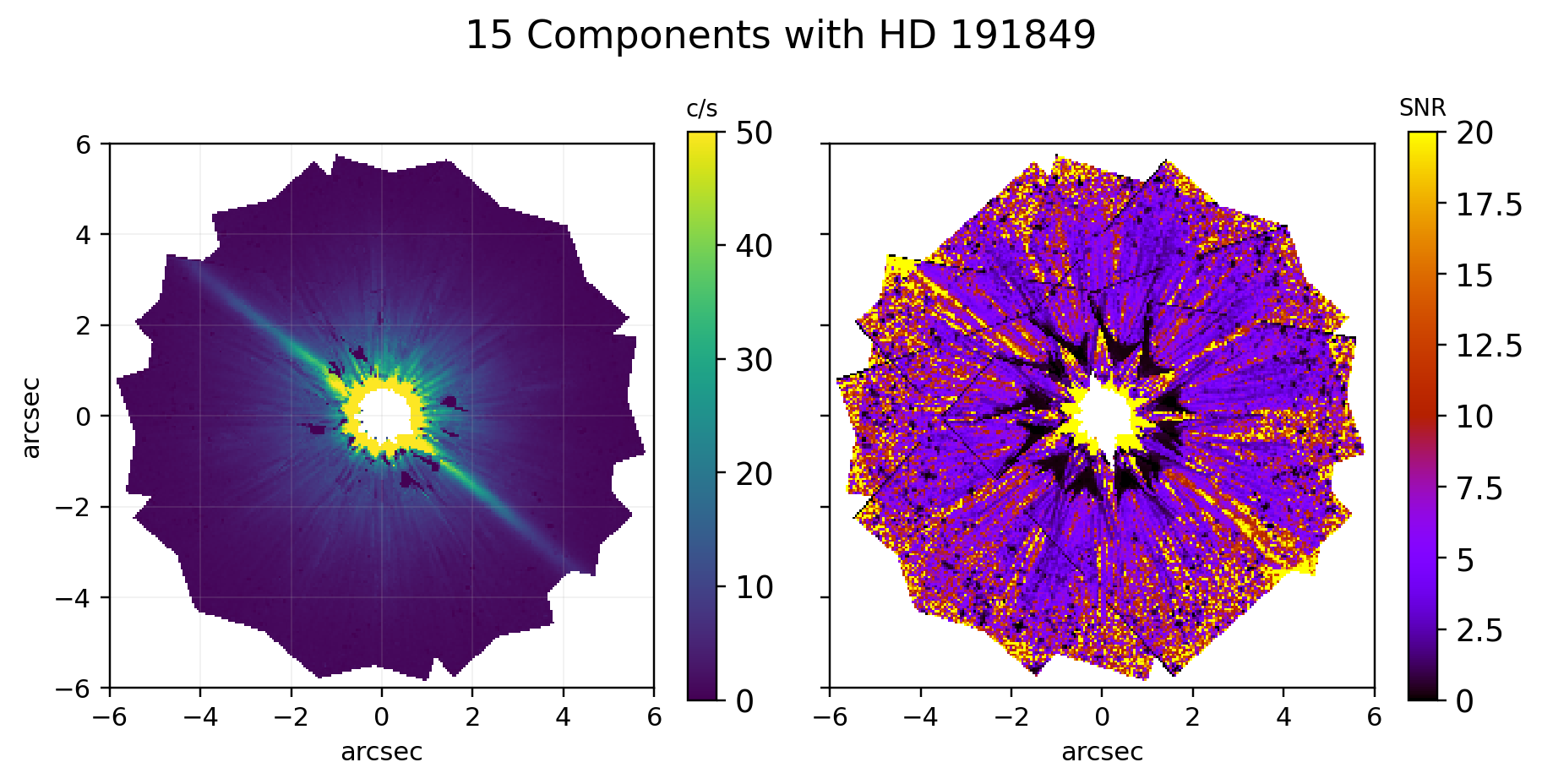}\hfill
\includegraphics[width=0.5\textwidth]{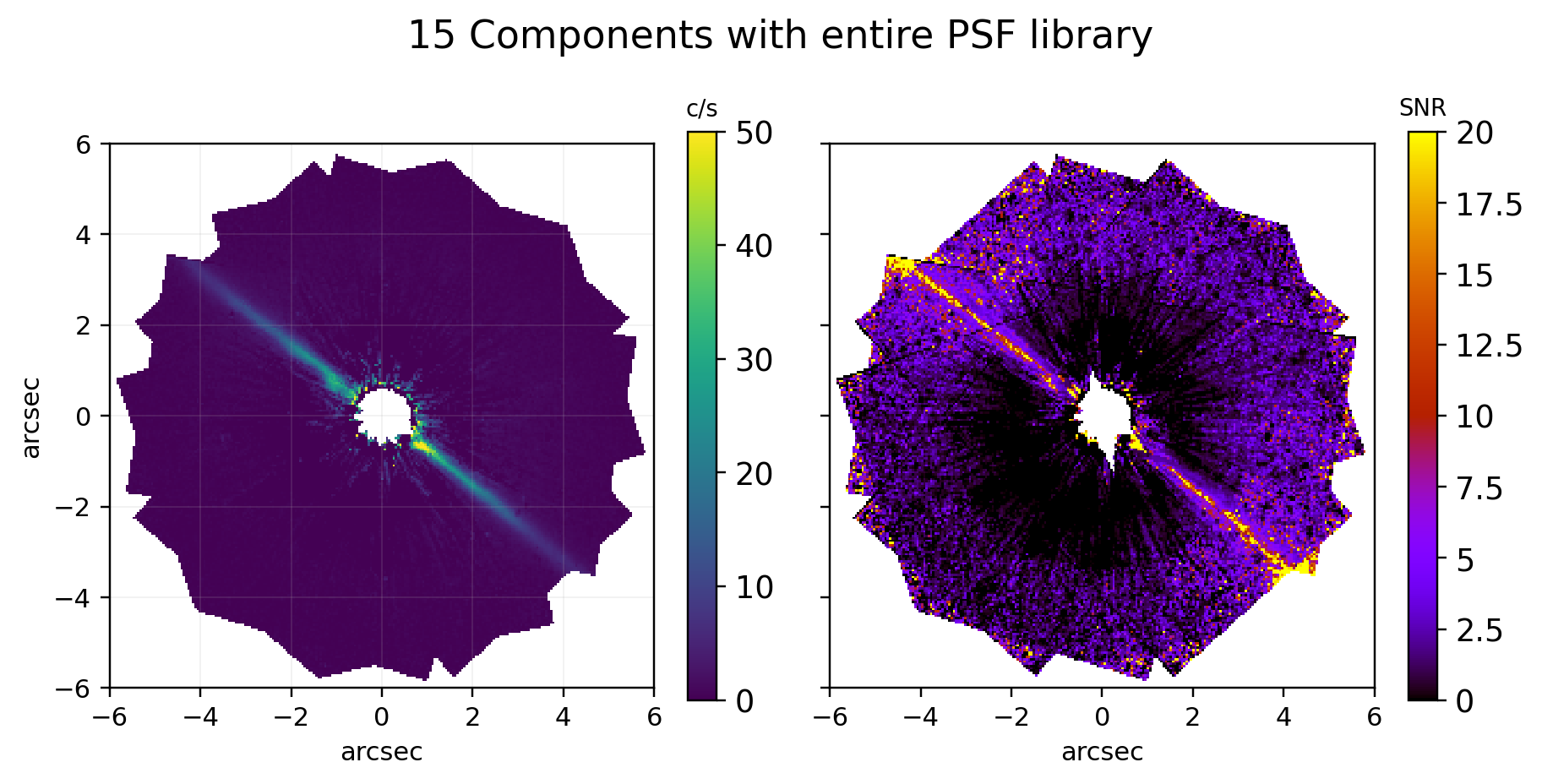}
\includegraphics[width=0.5\textwidth]{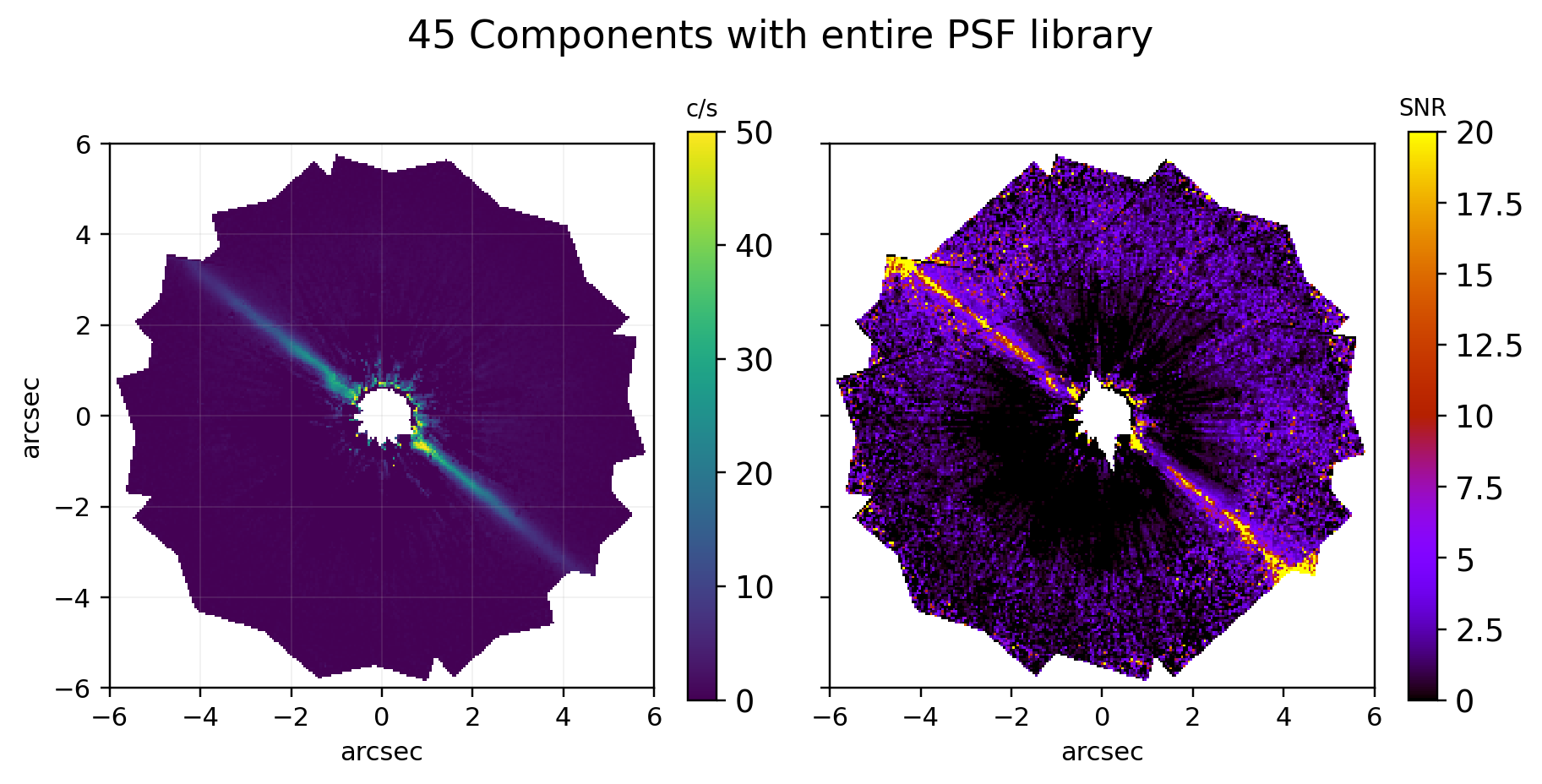}\hfill
\includegraphics[width=0.5\textwidth]{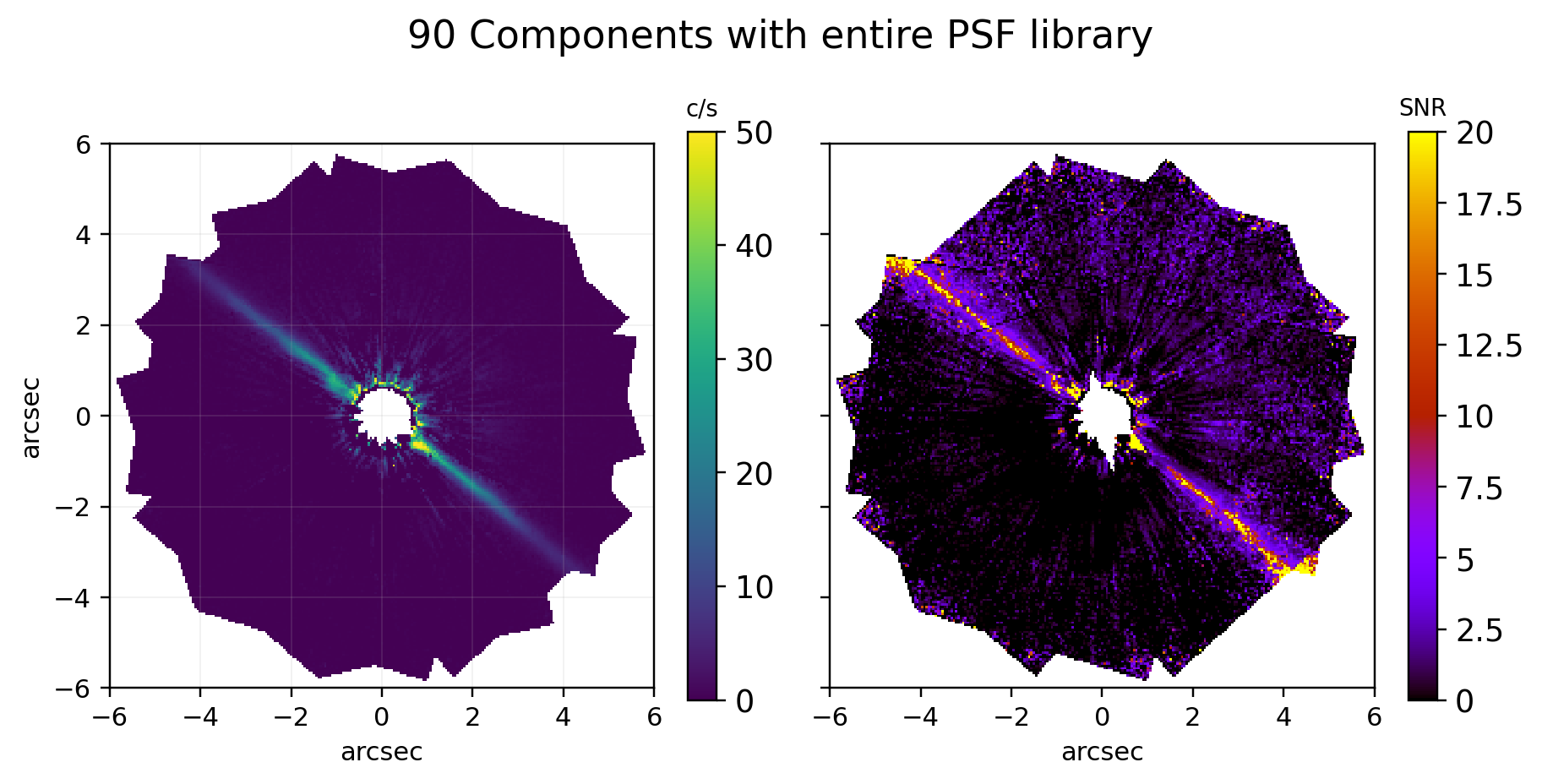}
\caption{\textit{Top left:} NMF reduction of AU Mic with 15 components and HD 191849 as a reference star. \textit{Top right:} NMF reduction of AU Mic with 15 components utilizing a PSF library. \textit{Bottom left:} NMF reduction of AU Mic with 45 components utilizing a PSF library. \textit{Bottom right:} NMF reduction of AU Mic with 90 components utilizing a PSF library. An improvement in S/N is observed when using a PSF library as opposed to a single reference star.} \label{fig:AUMicimprovements}
\end{figure}

\begin{figure}[H]
\centering
\includegraphics[width=0.78\textwidth]{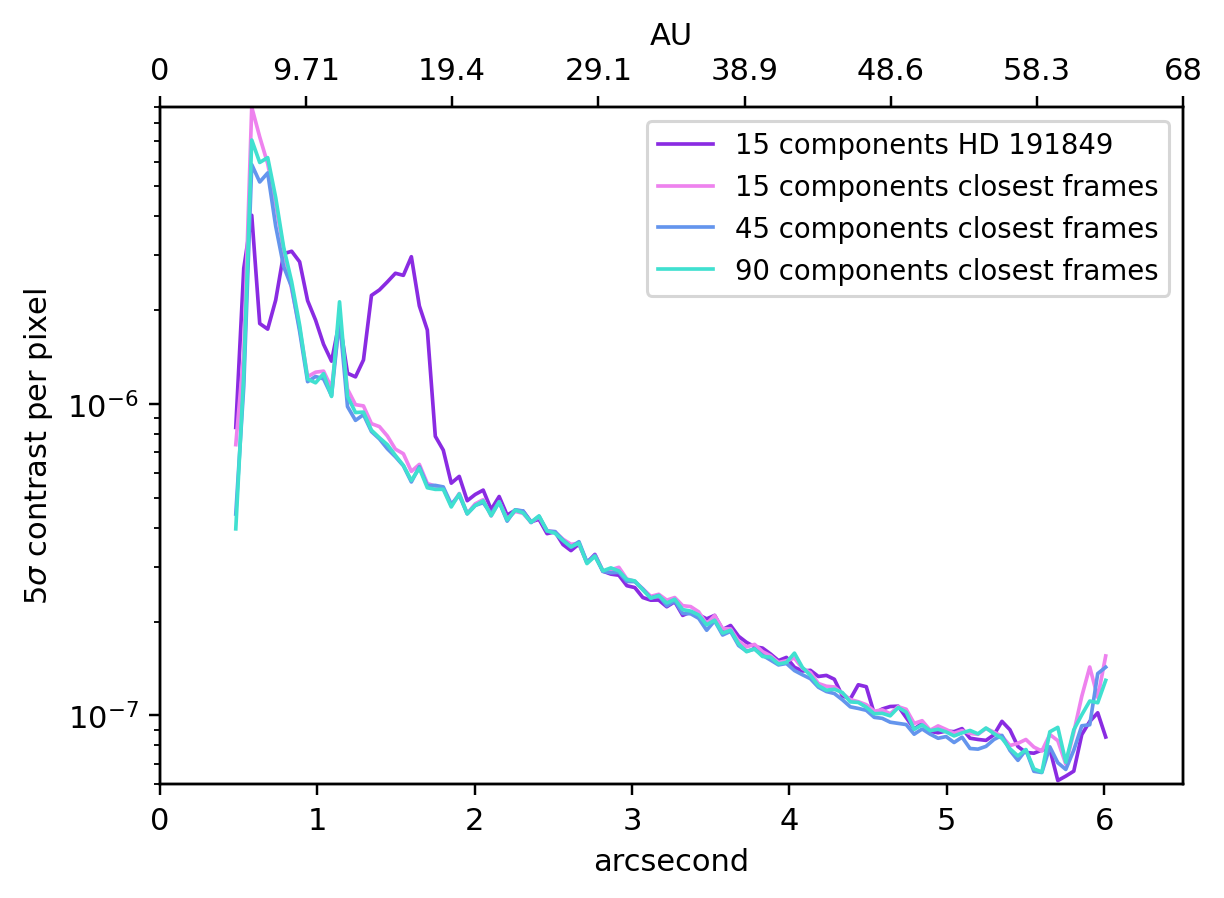}
\caption{5$\sigma$ contrast curves for the reductions shown in Figure \ref{fig:AUMicimprovements}. At closer separations, we note an improvement in contrast for the PSF library reductions compared to the single PSF reference star reduction.} \label{fig:AUMic Contrast}
\end{figure}

We note that the largest improvement to SNR comes from using the closest frames across the entire PSF library instead of a single PSF reference star. This is because the STIS coronagraph is sensitive to target and reference star color ($\Delta(B-V)$) mismatch, leading to rotationally-invariant chromatic residuals and causing under-subtraction of the disk as noted in Schneider et al.\cite{2012AAS...21944602S}. Following the selection of the closest Euclidean frames, we also see further SNR improvements with increased NMF components. To demonstrate this, we compare a reduction of AU Mic using a single reference star and NMF reductions of varying numbers of components using a PSF library of all PSF stars in the same occulter position (Figure \ref{fig:AUMicimprovements}). We note a decrease in the number of speckles both near and far from the central occulted region. We also note improvements in SNR across the disk surface. In Figure \ref{fig:AUMic Contrast}, we show a significant improvement in contrast by selecting the closest Euclidean frames compared to a single PSF reference star, while a higher number of components either provide a modest increase in contrast or no change. 

\subsection{Improved Efficiency with a GPU} \label{sec:gpu_efficiency}

\begin{figure}[H]
\centering
\includegraphics[width=0.8\textwidth]{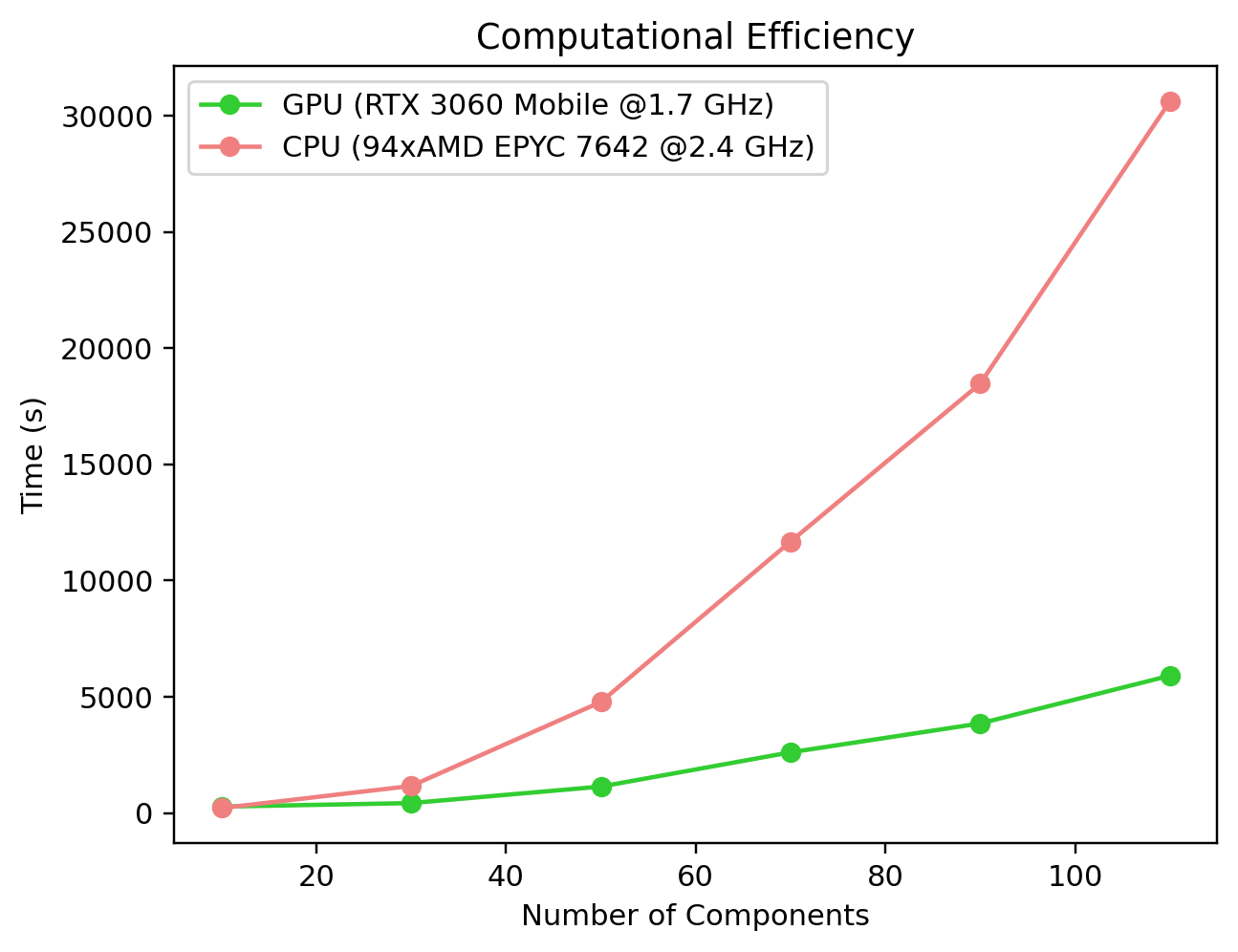}
\caption{Computing wall time as a function of a number of NMF components for a CPU configuration and a GPU configuration.} \label{fig:GPUvsCPU}
\end{figure}

With the large difference in thread count for GPUs compared to CPUs, we expected to see significant run-time improvements. We tested efficiency by performing reductions with differing numbers of components with a CPU setup and a GPU setup. We demonstrate in both Figure \ref{fig:GPUvsCPU} and Table \ref{tab:gpu_efficiency} the improvement in computational efficiency with an increase in the number of NMF components. We note that the CPU is faster than the GPU for the smallest number of components (10). This is because of the memory transfer overhead, which increases linearly, from transferring arrays in the CPU to undergoing computations on the GPU. As we increase the number of components, the overall matrix computation optimization from the GPU dominates over data throughput overhead. 

\begin{table}[ht]
\label{tab:gpu_efficiency}
\begin{center}       
\begin{tabular}{|l|l|l|l|} 
\hline
\rule[-1ex]{0pt}{3.5ex}  Number of components & CPU wall time (s) & GPU wall time (s) & Efficiency Increase \\
\hline
\rule[-1ex]{0pt}{3.5ex}  10  & 210   & 273  & 0.77 \\
\hline
\rule[-1ex]{0pt}{3.5ex}  30  & 1162  & 425  & 2.73   \\
\hline
\rule[-1ex]{0pt}{3.5ex}  50  & 4776  & 1135 & 4.21   \\
\hline
\rule[-1ex]{0pt}{3.5ex}  70  & 11676 & 2609 & 4.48  \\
\hline
\rule[-1ex]{0pt}{3.5ex}  90  & 18471 & 3852 & 4.80  \\
\hline 
\rule[-1ex]{0pt}{3.5ex}  110 & 30610 & 5907 & 5.18  \\
\hline 
\end{tabular}
\end{center}
\caption{Improvement in computational efficiency using a GPU instead of CPU with an increase in the number of NMF components. The CPU is faster for 10 components because of the overhead of transferring data from the CPU to the GPU.} 
\end{table}

\section{Summary and Future work} \label{sec:summary}

We have shown five different disk reductions in this paper, shown in Figures \ref{fig:NMF disk reductions} and \ref{fig:AUMicimprovements}. In the future, we will use our pipeline on other archival datasets to recover disks and conduct scientific analyses, such as fitting disk models and potentially identifying additional structures. Additionally, as our initial result for HIP 65426b seems promising, we plan to investigate the efficacy of NMF in recovering planets/companions compared to other existing post-processing methods, such as KLIP and LOCI. We have shown that higher numbers of NMF components coupled with our pipeline better recover disks and have improved contrast. We have also demonstrated greatly increased computational efficiency of GPUs compared to CPUs, particularly for higher numbers of components. We plan to explore alternate implementations of NMF, which could help with even better computational efficiency and/or better convergence properties.  

\acknowledgments 

This research is based on observations made with the NASA/ESA Hubble Space Telescope obtained from the Space Telescope Science Institute, which is operated by the Association of Universities for Research in Astronomy, Inc., under NASA contract NAS 5–26555. These observations are associated with programs HST GO-12228, HST GO-152183, and HST GO-15437. 

This work is also based on observations made with the NASA/ESA/CSA James Webb Space Telescope. The data were obtained from the Mikulski Archive for Space Telescopes at the Space Telescope Science Institute, which is operated by the Association of Universities for Research in Astronomy, Inc., under NASA contract NAS 5-03127 for JWST. The authors acknowledge the team led by PI Sasha Hinkley for developing their observing program, ERS 1386, with a zero-exclusive-access period. 

Portions of this research were supported by funding from the Technology Research Initiative Fund (TRIF) of the Arizona Board of Regents and by generous anonymous philanthropic donations to the Steward Observatory of the College of Science at the University of Arizona. 

This research made use of \texttt{astropy}\cite{2022ApJ...935..167A}, \texttt{ccdproc} \cite{matt_craig_2017_1069648}, \texttt{cupy}\cite{cupy_learningsys2017},
\texttt{matplotlib}\cite{Hunter:2007},
\texttt{multiprocess}\cite{McKerns_Aivazis_2010}\cite{mckerns2012building},
\texttt{numpy}\cite{harris2020array},
\texttt{pandas}\cite{reback2020pandas},
\texttt{pyKLIP}\cite{2015ascl.soft06001W},
\texttt{radonCenter}\cite{bin_ren_2018_2399553},
\texttt{scipy}\cite{2020SciPy-NMeth}, and
\texttt{SpaceKLIP} \cite{2022SPIE12180E..3NK}.

\bibliography{report} 
\bibliographystyle{spiebib} 

\end{document}